# How substance-based ontologies for gravity can be productive: A case study

Ayush Gupta[1], Andrew Elby[1,2], Luke D. Conlin[3]
[1]Department of Physics, University of Maryland, College Park, MD 20742.
[2]Department of Teaching and Learning, Policy and Leadership, University of Maryland, College Park, MD 20742.
[3]Graduate School of Education, Stanford University, Stanford, CA 94305.

Many science education researchers have argued that learners' commitment to a substance (matter-based) ontology impedes the learning of scientific concepts that scientists typically conceptualize as processes or interactions, such as such as force, electric current, and heat. By this account, students' tendency to classify these entities as substances or properties of substances leads to robust misconceptions, and instruction should steer novices away from substance-based reasoning. We argue that substance-based reasoning, when it supports learners' sense-making, can form the seeds for sophisticated understanding of these very same physics concepts. We present a case study of a group of elementary school science teachers in our professional development program. The teachers build a sophisticated explanation for why objects of different masses have the same acceleration due to gravity, starting from substance-based metaphors for gravity. We argue that, for conceptual, epistemological, and affective reasons, instructional interventions should focus on tapping these productive matter-based resources rather than attempting to replace them.

## I. INTRODUCTION

Research in physics and science education has documented the existence and persistence of naïve ideas held by learners in domains such as mechanics, thermodynamics, and electricity and magnetism [1–6]. Researchers generally agree that, in the process of learning correct science concepts, learners must articulate and in some cases rethink their naïve conceptions [7,8]. Researchers often disagree, however, about when and how instructors should guide students during that process. These disagreements stem in part from different accounts of the cognitive structure of students' naïve ideas [9].

For instance, some researchers attribute comparatively robust, stable "misconceptions" to students,

misconceptions that cannot contribute to a more expert understanding [1,2,10]. By this account, teachers should help students confront their misconceptions early in the conceptual change process, so that students do not incorporate the misconception into the new knowledge structures they construct. By contrast, other researchers describe student's naïve conceptions as consisting of loose collection of abstractions based on physical experiences in the world, such as *closer is stronger* and *more effort produces more result*, and incorrect ideas as resulting from the contextual activations of these knowledge fragments [11–14]. By this account, conceptual change consists largely of restructuring those intuitive ideas, and hence teachers should look for and help students build upon the seeds of productive thinking in students' reasoning, even in their incorrect conceptions [9,15].

Other models of cognitive structures and conceptual change fall somewhere in between. Some such models assume that naïve knowledge and cognitive dynamics are strongly constrained by particular cognitive commitments [16,17] but allow for variability in students' ideas (and underlying cognitive dynamics) that do not violate those commitments. A prominent line of research within this category argues that novice commitments are *ontological* in nature, i.e., that students have robust preconceptions about the nature of the physical entities they encounter in science courses [17–22]. Specifically, according to this framework, novices incorrectly persist in conceiving of scientific *processes* (such as heat, force, and light) as *substances*; and these mis-ontologies lead to many of the specific, robust misconceptions documented in the literature. In keeping with this theoretical perspective, instructional reforms have been designed in introductory physics topics to steer students away from "incorrect" *substance* ontology and towards "correct" *process* ontology [21,23]

In this paper we challenge this characterization of *substance* ontology as unproductive, arguing that "incorrect" ontological reasoning can be productive for the generation of sophisticated and correct scientific argument and reasoning. In the case of some concepts such as energy, our argument is not particularly striking; when done judiciously, thinking of energy as analogous to conserved "stuff" can be useful in both novice and expert reasoning [24–27]. Our argument is surprising because we apply it to force, a concept that experts are less likely to talk about in terms of substance-based analogies, perhaps in part because force is generally not a conserved quantity. Despite the seeming unproductiveness of a *substance* ontology of force, we argue that thinking of gravity as "stuff-like" contributed to learners' conceptual process in learning about gravity, forces, and motions, progress that would likely have been less transformative for the participants had the instructors steered students away from this "mis-ontology."

We begin with a brief review of the literature on novice ontologies in physics in Section II. Section III then discusses the methodology of our data collection and analysis. Section IV presents our analysis of the data. We end with discussions and conclusions in Section V.

## II. SUMMARY OF RESEARCH ON NOVICE ONTOLOGIES IN PHYSICS

In this section we review the research on novices' ontologies in physics and the instructional implications. First we present the view that many novice misconceptions in science stem from students' mis-ontologies, specifically their incorrect, un-productive classification of certain entities as substances. Then, we present research challenging this view. We end by discussing the limited research on novices' ontologies of forces and the purported implications for instructors who encounter students using substance-based metaphors/descriptions for forces.

### A. Some researchers argue for suppressing substance-based ontologies for science concepts

In a prominent line of work, Chi and colleagues have proposed that novices' ontology of science concepts such as force, electric current, and light are the primary barrier to their learning of these concepts [17,18,20,28]. Here, a learner's *ontology* of an entity refers to her sense of the basic nature of that entity. For example, an entity such as table, could fit into many categories – *furniture, something made of wood or metal, something having four legs, an object in the dining room*, etc. – but at the broadest level most of us would consider a table to fit within the category of *objects*; in other words, *object* is the ontological category for table. The term ontology in philosophy is often used to refer to describe the nature of the world, the most basic categories in the world. In cognitive science and learning sciences the term has been used to refer to a psychological construct: the mental categories into which an individual divides the world [29,30]. It is in this latter sense that we use the term in this paper.

Chi and colleagues use this notion of ontology to help explain the expert-novice divide in science [17,18,20,28]. In this view: (1) many notoriously difficult-to-learn science ideas such as electric current, forces, heat, and light belong to the ontological category of *process*; (2) experts categorize these science ideas correctly as "processes," while (3) novices tend to misclassify these ideas as *substances/matter*; and additionally, (4) novices are committed to the incorrect categorization. Thus, mis-categorization and resistance to re-categorization leads to robust misconceptions about these entities. In this view, the first step in successfully teaching these topics must involve interventions that help students replace their mistaken ontological categorization with the correct one. During and after this intervention, the instructor must carefully avoids any language, analogy, or metaphor that might cue or reinforce the incorrect ontological category.

This view has been taken up by other researchers and has influenced theory-building on conceptual change as well as instruction in secondary and college classrooms. [23,31–40]. For instance, Brookes and Etkina [41–43] have proposed that experts often use metaphorical language that encodes incorrect ontologies, for example using the metaphor of a physical well to describe a potential function in quantum mechanics. Novices tend to unproductively adopt and extend the ontologies embedded in experts' linguistic expressions, leading to incorrect responses or ideas which can become resistant to instruction [41,43]. In earlier work, Brookes [44] proposes that students often categorize science concepts when first introduced to them, and so, in order to avoid miscategorizations, an instructor should avoid any language and representations that encodes incorrect ontologies in its grammatical structure. In later work, however, Brookes and Etkina [42] have moved away from this recommendation (see the next sub-section).

## B. Some recent work argues that *substance* ontology can be productive for learning physics

Other researchers have argued *against* the idea that substance/matter based ontologies are un-productive. We discuss these arguments in brief and connect them to the particular argument we are making in this paper.

diSessa has argued that expert understanding of many non-matter concepts such as energy and momentum involves matter-based cognitive elements such as resources for understanding conservation.

Gupta, Hammer, and Redish [45,46] argue that both experts and novices flexibly shift in their ontological stances toward physics concepts such as light, electric current, and heat. They present a case study of a classroom discussion. In one segment of the discussion, a student, Kimberly, talks about electric current as a substance moving in the wires. Later in the discussion, she shifts to talking about current as a process in which electrons move in the wires. Kimberly's conceptualization of matter-like entity (what she called "current") moving through the wires was productive in her generating a correct understanding of Kirchhoff's junction rule in parallel circuits. Thus, Gupta *et al.* [45] argue that substance-based ideas are not a barrier to generating correct understanding.

Within physics education, researchers are increasingly attending to the ontologies of specific science concepts, their use by novices, and the instructional affordances and constraints of particular ontological descriptions [24,25,27,47–56]. Swackhamer [24], Brewe [25] and Scherr [26,27] have argued for the instructional productivity of a substance ontology of energy, conceptualizing energy in terms of everyday ideas used for material objects such as location, storage, conservation, and movement, but without necessarily burdening it with every single attribute of a material object (such as having weight or color). Brewe [25] shows how students can productively use the energy-as-a-substance metaphor for problem solving in physics. Brookes and Etkina [41,42] have argued that experts often use ontological metaphors, e.g., productively using matter-based ideas for non-matter entities. Expressions such as "particle in a well," "tunneling," and "potential well" draw on matter-based conceptions for non-matter concepts such as the potential due to an electric charge and the quantum motion of a particle in that potential. They have argued that language use that is often diagnosed as reflecting a misconception might actually be a symptom of students' struggling to understand the ontology of a science concept and that the appropriate instructional response is to attend closely to students' use of ontological metaphors to help students make sense of the ideas: "physicists' language indicates that they categorize the concept of 'force' into different ontological categories, depending on context. In the case of ontological disambiguation, we, the teachers, may simply be failing to hear the questions our students are asking. Namely, students need help sorting out the different ways in which the concept of force is categorized ontologically rather than help being disabused of certain mistaken ideas they might have." ( [42], p. 010110-11). Baily and Finkelstein [47] explore whether and how students draw on quantum versus classical ideas (and associated wave versus particle ontologies) to explain phenomenon such as electron diffraction and interference.

Having briefly reviewed the debate about the productivity and unproductivity of novices' use of *substance* ontology, we now turn to some research on the ontologies underlying novices' conceptions of forces.

### C. Research on novices' ontologies of physical forces

Research on novices' conceptions of basic mechanics is extensive [3,4,6,57,58]. Much of this work does not directly refer to novices' ontology of force concept; but some of the research can be and has been re-conceptualized in terms of ontology. For example, Brown [59] claimed that students' incorrect responses on Newton's third law questions arise because they conceptualize force as the property of an object. Reiner et al. [19] draw from a variety of research on students' understanding of mechanics to argue that novices think about physical forces in terms of a material substances schema: "Thus, force can be seen as a process of interaction between two objects … All these concepts are misconceived by novices as material substances or at least as possessing attributes of material substances." ( [19], p. 11). Reiner et al. [19] argue that documented novice misconceptions in mechanics such as associating a higher force with a faster object, conceptualizing gravity as the potential for all objects to fall, or the notions that force can be transferred from one object to another and can dissipate reflect a substance ontology of forces. Drawing on an historical analysis by Jammer [60], Brookes and Etkina [42] classify the historical use of the term force by scientists as one of four ontological metaphors: as an agent, as an internal drive or desire, as a passive medium of interaction, or as the property of an object. They argue that students often view forces as a property of motion, rather than of an object; and that their difficulties likely stem from trying to disentangle the different ontological metaphors available for talking about forces: an act of "ontological disambiguation."

Mostly, researchers have regarded students' use of forces as a property of objects as problematic. Researchers who argue for the productivity of *substance*-based ontological metaphors/reasoning about non-material entities have discussed entities such as energy, momentum, and fields, but not forces. (One exception to may be Brookes and Etkina [42], who note that expert physicists use language that could reflect forces having agency, for example. While they do not explicitly take a stance on the productivity of a substance ontology for forces, their observation that expert physicists use such language could suggest that they do not consider such use to be unproductive in the case of novices).

In light of the unsettled debate on the appropriate instructional response to learners' use of incorrect ontologies, what might be a proper instructional response when novices draw on *substance* ideas for forces? In this paper, we contribute to this debate on instructional strategies by analyzing a group of elementary- and middle- school teachers (physics novices) trying to figure out why heavier and lighter objects fall at the same rate when air resistance is negligible. At key moments, their language and reasoning reflects an underlying ontology of gravity as a material agent/substance. Our analysis bolsters the argument that *at least in some instances* substance-based metaphors for non-material entities—even forces—can be productive for physics learners. Therefore, instruction should not necessarily aim to immediately push students away from using such ontological "miscategorizations." Indeed, interventions that prematurely curb either linguistic or conceptual use of incorrect ontologies might be detrimental to learners' progress.

# III. DATA COLLECTION AND METHODS OF ANALYSIS

## A. Setting

Our analysis focuses on a discussion on falling objects by middle and elementary school science teachers in a two-week summer professional development workshop in 2009. The workshop aimed to engage the teachers in scientific inquiry and to assist them in developing resources for facilitating inquiry-based science instruction in their own classrooms. 18 teachers participated, facilitated by 6 researchers (including the three co-authors). During the first week, teachers engaged for three hours per day in an extended inquiry on the motion of falling objects. The discussions were largely driven by the ideas and explanations generated by the teacher participants. They cycled through small groups discussions (groups of 4-5) and whole-group discussions. The lead facilitator coordinated the transitions between small and large group discussions. Small-group discussions allowed participants to generate or refine ideas in a setting that afforded greater focus and a safer space for the quieter participants to contribute. Typically, one of the facilitators would listen in on a small-group discussion, moderating as needed. Whole-group discussions served to share and/or contrast ideas and explanations, to make progress toward consensus, evaluate emerging evidence, or decide on next steps. All small- and large-group discussions were videotaped.

## B. Data Selection

At the end of each workshop day, the facilitators debriefed to share their impressions of the inquiry session and note particular moments in small-group and large-group discussions that might contribute to the project's research questions, which included how teachers come to participate in authentic scientific inquiry.

At the end of the workshop, the researchers compiled a list of notable moments from the workshop, which were then transcribed. These transcribed video-records were then analyzed in video analysis meetings in which the entire research team participated. The transcript in this paper follows the typical conventions in physics education research literature. The only things to note are: (i) ellipses (…) denote a pause in the speech, (ii) words surrounded by slashes (/) represent our best attempt at capturing utterances that were difficult to hear, and (iii) italicized phrases in regular parenthesis () indicate gestures or other non-verbal information. Punctuation in the transcription relies on the prosody (pitch, rhythm, intonation) and on the conventional interpretation of the utterance.

In this paper, we present two related episodes in which the participants discuss why objects of different masses fall with the same acceleration. The first episode is a small-group discussion. The second episode comes from the ensuing whole-group discussion moderated by Elby.

We selected these episodes for the following reasons:

1. Episode 1 (Small Group): The teachers in the selected group authentically struggled to *explain* why objects of different weights "fall at the same rate," a result they knew from before. In doing so, the teachers described gravity using multiple analogies and metaphors that provided rich insight into the ontologies that the teachers attributed to the concept of gravity. In our initial viewing of the videotape, we noticed that some of their language and analogies for gravity seemed to reflect a substance-based ontology. Since we knew that these teachers had come up with interesting conceptual ideas, we decided to explore whether and how their use of substance-based language and analogies affected their conceptual progress in answering this question.

2. Episode 2 (Whole Group): In this episode, the teachers from the small group share their ideas with the whole group. We selected this episode because (i) in our preliminary viewing we saw evidence of ontological variability , (ii) we were struck by the "coolness" of the reasoning Lynn articulated for the whole group, and (iii) we wanted to analyze and showcase how they made progress toward constructing a correct understanding of why objects of different weights have the same acceleration when falling.

As shown below, it turned out that some of the ontologies the teachers attributed to gravity were "incorrect." Given the debate in the literature on whether incorrect ontological reasoning should be quickly challenged by the instructor [46,61] in order for correct conceptions to develop, the selected episodes had the potential to inform this debate.

### C. Analysis

Through a close analysis of the transcript of the small- and whole-group discussions, we characterize the ontology of gravity implied by the verbal utterances of the speakers. We rely primarily on two methods for inferring the ontological commitments in speech—predicate analysis [20] and analysis of grammatical metaphors [42]—each of which will be discussed in more detail below. At times, our participants made explicit analogies to represent gravity. We will also characterize the ontological commitments implied by these analogies. Although predicate analysis and grammatical analysis are distinct approaches, for our purposes we take them to be largely complementary. Both methods analyze lexical units to infer the implicit ontological commitments of speakers, and both draw similar distinctions between the ontological categories (e.g., *material substances* vs. *processes*). Although these approaches differ in the details of how they go about coding statements for these categories, both methods should by and large agree on the code for any given statement. In what follows, we will describe how each coding scheme works, and how we combine them.

### *1. Method of Predicate Analysis*

Predicate analysis, originally developed by Sommers [62], has been used for classifying the implicit ontological commitments in language [29]. The method is based on the observation that in natural language, a given target entity (e.g., *the tree*) cannot be paired with just any predicate (e.g., *is an hour long*). When a predication is not allowable, it signals that the ontological category of the entity (e.g., an *object*) is not compatible that of its predicate (e.g., a *process*). When the predication is allowable, the ontological category of the predicate can be used to infer the ontological category of the target entity.

Keil [29] applied this method to the language used by children and adults in order to study the interrelated psychological development of ontology, language, and conceptual understanding. Chi and her colleagues further adapted this method to explore students' ontologies of science concepts [17,18,20–22,28]. They draw ontological distinctions between three fundamental categories: *substance/matter*, *process*, and *mental states*. They argue that science concepts such as heat, light, force, etc. correctly belong to the category *process*, and that novices tend to incorrectly categorize them, say, as *matter*.

To evaluate students' ontologies of science concepts, Chi, Slotta and colleagues [17,20,28] have developed taxonomies of various predicates for both the *matter* and *process* categories, which they use to code transcripts of students' speech. For example, matter (but not processes) can *move* and can *consume* (or *be consumed*), and so *move* and *consume* are classified as substance/matter predicates that are often exemplified in speech. Consider the following hypothetical student's reasoning for how a light bulb lights when connected to a battery: "The current comes out of the battery, travels down the wire, and is used up in the bulb." Predicate analysis would code for the following predicates applied to electric current: "comes out," "travels down," and "is used up." Within Chi's framework, these predicates belong to the *matter/substance* category and this implies that the student has a *matter/substance* ontology of electric current [17,20].

### *2. Analysis of Grammatical Metaphors*

Like Chi, Slotta, & colleagues, Brookes and Etkina [42] use a close examination of language-in-use to infer students' ontology of physics concepts. In particular, Brookes and Etkina focus their analysis on students' ontology of *force*. Instead of predicates, their method focuses on the speaker's use of grammatical metaphors, in which a term grammatically functions in ways that may differ from its literal meaning. For example, physicists would generally agree that forces are interactions, not objects. And yet, physicists often speak of forces in ways that have them grammatically functioning as one of the objects participating in an interaction (e.g., "a force acts on the cart"), rather than *being* the interaction in their own right.

This method focuses on grammatical structures to infer ontological substance of talk, with noun participants implying a *matter/substance* ontology and verbs indicating *processes*. Drawing on a historical analysis of the concept of force in physics, they draw further distinctions between several substance-based metaphors that differ with respect to the role the noun participant is playing (active/passive) and its location (internal/external). In the phrase "a force acts on the cart", force is a noun participant that is playing an active role (the force is doing something to another object), and by acting *on* the cart it is implied to be located external to the cart. This corresponds to the *force as agent* metaphor. The other substance-based grammatical metaphors are *force as internal desire* (location: internal; role: active)*, force as passive medium of interaction* (location: external; role: passive)*,* and *force as a property of an object* (location: internal; role: passive).

### *3. Combining Methods*

For our purposes, the methods of predicate analysis and grammatical metaphors are generally complementary. When they overlap in application to our data, they generally agree. However, they do not always both straightforwardly apply to any given statement. For an example from our data, when a teacher says, "gravity's still pulling the heavier object down," gravity is playing the role of a noun participant, and so it is clear that Brookes and Etkina would code this as a substance/matter ontological metaphor—specifically, *gravity as agent* metaphor since gravity is playing an active role (pulling the object) and the phrase suggests it is located external to the object. It is not so straightforward to determine how Chi and colleagues would code this however, since "pulling" is not one of the predicates listed in their taxonomy of substance/matter predicates. This is probably just because their taxonomies were developed to handle students' ontologies of heat, light, and current, rather than force specifically, even though they fully expect their methods to handle the case of force as well [19]. We could guess they would code this teacher's ontology of force as substance, since "pulling" seems to predicate a substantive "puller." However, rather than guess how to extend the methods and taxonomies to deal with all of the predicates in our data, we will take a more conservative approach by relying more heavily on whichever of the two complementary methods gives a clearer signal in the case of any given utterance.

And while we draw on the predicate analysis and grammatical analysis methods, we do not completely subscribe to the theoretical assumptions underlying these methods. For example, we disagree with the presumption that novices' (and experts') language use generally implies they have stable, inflexible ontological commitments, an assumption underlying the predicate analysis method (for further discussion see [45,46,61]). At the same time, we agree that, at times, predicates, analogies, and linguistic structures in an utterance could reflect the speaker's in-the-moment ontology. Given these caveats, we show our close analysis of utterances to illustrate how we are drawing evidence for our claims; and we present the transcript in full, so readers can draw their own conclusions and challenge ours. At opportune points in our analysis, we also discuss some of the nuanced differences between our theoretical and methodological inclinations and those of predicate analysis or grammatical analysis. We hope that this will give the readers a more situated understanding of these differences than would result from a decontextualized "philosophical" discussion.

In the next section we start with our close analysis of the two targeted episodes, attending to the grammatical role of the terms "gravity" and "force," the predicates attributed to those terms, and the use of grammatical metaphors and analogies.

## IV. EPISODE 1: TWO INCORRECT ONTOLOGICAL METAPHORS FOR GRAVITY

In this section, we make the case that teachers in a small-group discussion drew on substance-based ontologies for understanding gravity, and that doing so was productive for them. We start by briefly recounting the events leading up to the small-group discussion.

### A. The lead up to Episode 1

The workshop started with the following question:

> *Suppose you are walking at a steady, fast pace in a straight path with your keys in an outstretched hand. If you wanted to drop (not throw) your keys into a trash can located ahead of you, should you drop the keys a little before you reach the trash can, right on top of it, or a little after you pass the trash can?*

Initially, there was no consensus. Some teachers said the keys would fall straight down due to the gravitational pull and so the keys should be dropped right above the trash can. Others said the keys should be dropped before reaching the can since they would keep moving forward after being dropped. A few said the keys should be dropped after passing the trash can,

*Gupta, Elby, Conlin: Substance Based Ontologies can be Productive*

May 5, 2013

emphasizing how something dropped from a moving vehicle seems to go backward due to the wind or how a handkerchief held out the window of a moving car streams backward, suggesting that it would move back if released. The teachers went on to do some trials of dropping keys as they walked in the corridor, and quickly came to consensus that when walking the keys would have to be dropped a little before passing the trash can. But this consensus led to further questions: What if they were running, or riding in a fast car? What if instead of keys you had a heavy object like a brick or anvil? What if you had a very light object like a feather? What if there was no gravity, as in outer space? What if there was no wind resistance, as on the moon?

By the second day, the starting question about keys dropping into the trash can had become a means to explore conceptions about the motion of falling objects, and trying to unpack the effects as well as mechanisms by which gravity and air resistance influence the motion of falling objects. The activities cycled among debating ideas, designing experiments to test some of the ideas, and carrying out those experiments—such as dropping a bottle filled with pennies and an empty bottle out of a moving car. The discussions and experiments focused teachers' attention on the trajectory of the falling object, how fast an object was dropping, how much time it took to land, and whether the time to land depended on the weight of the object and the speed of the car.

### B. Episode 1: The small group discussion

Within this unfolding discussion, our focal episode comes from day 5. The teachers were struggling with the observation that the bottle filled with pennies hit the ground at the same time as a lighter bottle filled with water. This reminded some teachers of the standard demonstration many had seen or taught, that a heavy and light object dropped from the same height land at the same time. Andy (Elby, second author) asks the teachers to discuss this issue in their small groups:

Andy: Um, so, talk, there's one proposal that how gravity works on something doesn't care about its weight, so talk about this in your small groups. We know a heavy and light object just dropped here on earth, without significant air resistance, they fall together. Why? What's the causal story for why?

We focus specifically on the discussion that ensues in the group of Lynn, Mira, Daniel, and Lisa. Lynn starts by recalling the experiment supposedly done by Galileo by dropping objects of differing mass from the Tower of Pisa:

Mira: What is that (*inaudible*)
Lynn: No matter what you drop, they fall the same...
Mira: So what, the paper just had the...
Lynn: air resistance, yeah if you discount air resistance, things, and there was that famous experiment they did off the Leaning Tower of Pisa or whatever, and they dropped the objects, or [XYZ] school where she drops the computer and the…and they all fall at the same time...Even though I know that it's going to fall at the same time, I feel that deep down in my heart, gravity's still pulling the heavier thing down…And I feel like it gets confusing with the whole idea of (*something inaudible about going* *forward, with a forward hand motion*) when you throw something with more force, you know, it goes farther, two objects regardless of weight, the one I threw with more force would go farther, but I think I'm having a hard time disassociating the idea that those things that we see from just pure effects of gravity…because /even when/ we add force and we add air resistance…I don't know what causes gravity…

Lynn's statement indicates that she knows that the dropped objects of different masses fall the same distance in the same time but what is striking is her metacognitive reflection about how she feels about that answer. "Deep down" she feels that her intuition is at odds with the answer. Her intuition is that gravity would pull harder on the heavier object. She relates this to her everyday intuition that more force would imply a greater distance for the same thrown object. In what follows, resolving this tension becomes the focus for this group. Lynn takes her intuition very seriously in spite of knowing the right answer and seems troubled at the inconsistency between the two. As shown below, this tendency to seek coherence between multiple ideas is shared by the other members in her small group and is a key epistemological resource that enables them to make significant progress towards resolving this tension.

Taken literally, Lynn's statement, "gravity's still pulling the heavier thing down" suggests that she is treating gravity as an *agent* capable of action ("pulling"). Grammatically, in this phrase, gravity is playing the role of an active noun *participant* that is external to the other interacting object ("heavier thing"), reflecting an *agent metaphor* within the substance/matter ontology [42]. For force, the phrase "throw something with more force" suggests "force" as an entity that accompanies either the object being thrown or the action (the throw), depending on whether we interpret "with more force" as an adjectival or an adverbial phrase. Grammatically, "force" in this utterance takes on the role of a *noun* participant in a passive role. But it is not clear whether it is simply exerted on "something" or whether it becomes a property of it. So, from a metaphorical analysis, the utterance could suggest one of two metaphors; it could suggest a "force as passive medium of interaction" metaphor or a "force as a property of object" but it is not unequivocally clear which one it is. Either way, this statement would be coded as reflecting a substance-based ontology of force.

*Brief Reflection:* To be clear: in this analysis, we are not taking a stance about whether Lynn's use of language reflects some stable psychological reality for her (e.g., she really thinks of forces and gravity as matter-like) or whether she is drawing on everyday language about force [64] that has such metaphors and metonymies woven in its fabric, often invisible to the user entrenched in that language [63]. Our point is simply that Lynn is using language that draws on properties typically attributed to objects/matter to describe her ideas about gravity and force.

Given that Lynn's language incorporates ideas that are not strictly "correct," what is the correct instructional move for a facilitator in this instance? Should the use of language that incorporates incorrect ontologies be allowed to continue or be nipped in the bud? Some researchers have suggested that allowing the use of incorrect ontologies (and language that incorporates or supports incorrect ontologies) can severely impede the development of correct understanding of concepts [18,20,21,44,61] and that such incorrect ontological utterances should be weeded out of the conversation at the earliest. In this paper we argue against this position, showing how incorrect ideas can contain the seeds of productive discussions and correct understandings. In the case of Lynn and her group, the incorrect and conflicting ontological metaphors for gravity become the source for generation of a sophisticated argument for why objects of different masses fall with the same acceleration. *End of Reflection.*

After Lynn's statement, Lisa tries to argue that gravity would adjust for the weight of the objects, but cannot articulate a mechanism. And Lynn struggles with trying to think about a heavier object as made up of many little objects:

Lisa: What was the question?
Lynn: How can we explain the fact that two things fall at the same rate, even though [their weight]?
Daniel: I mean...
Lisa: Because if gravity (*inaudible word*) it will accommodate for the weight... like no matter what it is, it will... in my mind it's like
Lynn: Here's a nickel .... (*drawing separate quarters*) if you think of a quarter as having the same weight as 5 nickels, so if you, if gravity has whatever pull it has, it pulls that down (*pointing to drawing on paper*) And then basically, a larger object, just has the same pull as each of those /parts of/ ...so that…
Here's a whole tootsie roll, it's divided into parts, it's the same per mass, it's the same /holes/, so the same /equal/, so...

This idea of conceptualizing a more massive object as a collection of lighter objects ("a quarter as having the same weight as 5 nickels") later became a central idea within their group, driving the discussion. Here, it is not entirely clear what Lynn means by "gravity has whatever pull it has." But, grammatical analysis again places gravity in the role of a noun *participant* (a *matter ontology*). Also note how "pull" in this sentence also takes on the role of a noun *participant* (linguistically serving as a property of gravity). Gravity acts as an active *agent* in the utterance "it pulls that down," taking "it" to refer to gravity.

Daniel takes Lynn's statement to mean that gravity would pull equally on the light and heavy object and questions that reasoning:

Daniel: But when they're connected, why wouldn't you have /more pull/
Lynn: well, because there's no...this part takes up this part of gravity, this part takes up, so it's like it works on mass, so that on one gram, it has a certain pull, but it just works on that gram, so if you add another gram to it, you don't double the pull of gravity on it, but it now has two grams to pull down.
Daniel: I think that's as good as anything we'll come up with, but the question I want to ask is why don't you add these together?

Lynn: Because you still have the same amount of gravity, you add mass, so it only has, it still needs that (*inaudible word*). Let's say it takes one gallon to move a 500 pound car, then it's going to take two gallons of gasoline to move a 1,000 pound car, so if you think of gravity as a force like that... even though there's more of something, that more of something has to have gravity working on it as well, kind of like gravity is a force.

Lynn's argument here conceives of a heavy object as made up of parts, each of which is lighter than the combined object; there is a certain gravitational pull on each part; and, when you add two light parts together gravity is not doubled. By thinking of a larger object as made up of smaller identical pieces and each piece as having a fixed pull associated with it, and mapping gravity onto the pull on the smaller piece and not onto the total weight of the object, it seems Lynn is trying to accommodate both (i) the answer she knows to be correct (that the objects "fall together") and (ii) her intuition that the object with larger mass feels heavier (i.e., pulled down with a greater force). Lynn's explanation here, though more elaborate than before, is still not reconciled with Daniel's argument for adding up the gravitational force on each part when putting multiple parts together. Note that in trying to reconcile her conflicting ideas, and addressing Daniel's concerns, Lynn has effectively mapped gravity onto the force per unit mass, which would be the correct definition for gravitational field (i.e., the acceleration due to gravity).

Here, Lynn's analogy in which gravity is mapped onto the amount of fuel in a car suggests an ontology of gravity as *matter/substance*. "Takes up," "part of" and "amount of" are all *matter/substance* predicates. The phrase "works on" suggests the attribution of *agency* to gravity. For Daniel, Lynn's insistence that the larger object has the same gravity as the smaller parts violates his intuition that they should "add" (a *matter* predicate) the gravity of the parts to get the gravity of the whole. The predicates "more" and "add" are *matter-based*. Grammatically, in all of the utterances, gravity plays the role of a *noun participant*, suggesting a matter-based ontology.

*Brief Reflection:* Part of the disagreement between Lynn and Daniel might stem from ambiguous use of pronouns. When Daniel poses the question, "why wouldn't you have more pull?" it is not clear what the "more pull" acts on. Daniel could have been thinking about the gravitational pull on the connected object as a whole. Lynn's response, on the other hand, suggests that she might be opposing the idea that each component *part* of a larger object will not experience greater gravitational pull: "if you add another gram to [the first gram], you don't double the pull of gravity on [the first gram]" While the extensive use of pronouns might have contributed to the lack of clarity, we see the ensuing argument as productive for the participants. Daniel's insistence at reconciling Lynn's idea of equal gravity with the argument that a larger object should have more pull pushes Lynn to articulate her idea better. This kind of argumentation is crucial to creating knowledge and is an authentic science practice [7,8]. The workshop facilitators, therefore, faced a tension between supporting this argumentation by letting it continue versus nudging the participants toward a correct ontology of gravity and force. In this instance, the facilitators let the conversation continue without interjection. *End of Reflection.*

Daniel continues to argue against the idea that we can add objects without adding up the force of gravity on them – partly to get a clear sense of Lynn's reasoning and partly to problematize it. Following Lynn's analogy of fuel in a car, Daniel also tries to quantify gravity, now through the example of dividing an object into two lighter parts:

Daniel: So wait a minute, you've got X gravity to pull this down (*holds up whiteboard eraser*) If you divide this in two, then the thing's likely to take one half X to pull this down, and one half X to pull that down (*pointing to each half of the eraser*)

Lynn: No, no, because now it is (*inaudible word*)… Say there's a force of gravity here. I can't come over and borrow that force. The pull here, where this is sitting, there's a constant amount of gravity in this section of it, it's going to pull this down, but for this part to come down, it needs to use this gravity over here. It's all along the earth, so each little section has its own little gravity

Mira: Like a force field!

Daniel: Okay, but I think that my question's not answered yet. This is a pound brick and it takes X amount of gravity to pull it down, you're saying there's still going to be X gravity pulling it down here (*pointing to one half*) and X gravity pulling it down here (*pointing to other half*), I want to know how it wouldn't be one half X and one half X.

Lynn: Well, think of it this way. If I cut this in half, and put one half over here and one half over here (*holding them apart*), they're still going to come down at the same rate, right?

Daniel: Mmm Hmm.

Lynn: So, just the fact that they're joined together doesn't mean that it's going to take up more gravity. If I cut this into little tiny pieces, each, you would argue that, like if we took that penny bottle, and we had all of the separate pennies lined up, they would still all come down at the same rate, but what you're arguing is the fact that we're putting them all together means they need less gravity to come down, and I guess what I'm arguing is that each mass, each little mass needs its part of gravity that pulls down, so the fact that you've added a bunch together doesn't mean that they're going to come down quicker. Each little section still needs the same pull of gravity to pull it down.

Here Daniel's counter-arguments make explicit his assumption of conserving gravity: the total pull on an object should be the sum of the pull on the parts.

In response to Daniel's objection, Lynn further fleshes out an argument for why we cannot add or divide the gravity when we think of adding two objects to make a heavier one or dividing an object into lighter parts. Here, she reasons that each part "use[s] the gravity" at its own location. Here, Lynn is conceptualizing gravity as a constant pull spatially distributed, so that each part (of a larger object) only experiences the pull at its own location. The falling of a heavier object, in Lynn's reasoning, is just as if a bunch of lighter objects were falling under the influence of gravity; each (lighter) part falls the same irrespective of whether it is a separate object or part of a heavier object. So, the bottle of pennies falls alongside a single penny, because each penny-in-the-bottle is falling alongside the single penny.

We should note that Lynn's reasoning, of mapping gravity to the pull per mass-unit but rejecting that the gravity on the pieces can be added, provides a way to make sense of the phenomenon, but does not directly address why the more massive object feels heavier but has the same acceleration as a less massive object.

Ontologically, Lynn and Daniel continue to reason about gravity as being quantifiable, having spatial location, and having agency (the ability to pull an object). A predicate analysis of Lynn's and Daniel's reasoning reveals multiple categories of substance attributes: *less*, *more*, and *amount of* as markers of the quantity predicate; *use* and *takes* as markers of the consume predicate; and *borrow* as a marker of the absorb predicate. The phrase "over here" attributes a location to gravity, which is part of the substance schema from Reiner et al. [19]. Daniel's utterance "X gravity pulling it down," reflects the metaphor of gravity as an agent. But Lynn's developing argument has further nuance: it potentially rejects a "pure" substance ontology of gravity by rejecting the idea of adding/dividing gravity, i.e., by rejecting an aspect of conservation. Daniel's arguments, on the other hand, include the idea of gravity as a conserved quantity that is added up to a larger whole or divided into smaller bits depending on how the object is integrated or divided.

Lynn's reasoning fails to convince Daniel. And Lisa tries to compare Lynn's idea of thinking about the gravity associated with each part of a heavier object with Lynn's earlier analogy to fuel in a car. Lynn tries to tie these ideas together saying that if you need a "gallon of gravity" for pulling down a 500lb car, then a second 500lb does not change the rate of falling because the added mass comes with additional gravity. In this analogy, gravity is mapped onto the fuel per car:

Daniel: Yeah, but I guess I'm not completely convinced.
Lynn: (*laughing*) yeah, I can see I'm not explaining it well. But now….
Daniel: I mean it's a good idea, but it's, I don't know, the objection is going to be.
Lisa: I thought it was going to be like, you were going with the whole /formula/ thing (Lynn: Mmm hmm.) So…
Lynn: It is, it's equal on each little section of it, No, because you're thinking of the (*inaudible word*). I'm not thinking of the quarter as one whole, I'm thinking of the quarter as five little separate parts, and each one of those little separate parts needs the same amount of gravity to pull it down. So that you don't gain anything, you don't make it any harder.
Lisa: This equals this? (*pointing*) It's not like it's its own gravity, it's like it equals, each part of this equals this? It's like the gasoline thing sounds different from what you were explaining earlier. Because if you have 1 gallon for 500 pounds, and 2 gallons for a thousand, it's like whatever the formula is, and it was good, but it's just like...

Lynn: Well, with the car, here's a gallon of gas, it's like a gallon of gravity. That can pull 500 pounds worth of car. But, if I add another 500 pounds of car, that doesn't mean it's all of a sudden going to come down faster, what it means it needs its share of gravity as well. So it's going to be constant, as I add weight to it, I'm actually using a little bit more gravity...

In this segment, Lynn continues to associate gravity with the substance predicate *amount*, suggested by the use of attributes such as *amount of*, *little bit more*, and *gallon of gravity*. She also uses the substance predicate *consume* via the analogy of gravity as consumable fuel ("using a little bit more gravity").

Lynn's final utterance suggests the notion of compensation, how the effect of adding more mass is compensated by using more gravity. (This notion of compensation becomes a vehicle through which the group later comes to a satisfying explanation.) Also, note Lynn's subtle shift away from her earlier stance that gravity is not add-able: here she says that when you add another car, "I'm actually using a little bit *more gravity*" (emphasis added). Following Lynn's explanation, Daniel tries to make sense of it. In doing so, he generates his own analogy for gravity. He analogizes gravity to student letter-grades; just as Lynn argued that each bit of mass comes with its own pull, Daniel says that each student comes with her own grade, so that adding a student does not raise or lower the grade that each gets. For instance, putting together three students, each with a grade of C, does not change their collective grade. He uses this to makes sense of Lynn's idea of how gravity could remain constant even as you add weight to an object:

Daniel: No, well you are /stipulating/ but I guess I'm just trying to think of what the objections will be.... I was trying to think of an analogy, and I haven't come up with a very good one, I don't think, you could think of it in terms of grading students on an absolute scale, so that, let's say I have two students, and they both can get the same grade, they get A's, they both can get A's. It's not like I only have a certain amount of A, so I have to divide that, so they both get C's, because there's not enough A-ness in order to give them both A's. So, if gravity, if that's a useful analogy, regardless of the /mass/ they still get A's.

Lynn: No, because I'm not saying there's a finite amount of gravity that has to be divided between everything, that's kind of what you're saying, that there's a certain amount of points. I'm thinking of gravity as something that's spread equally on the earth, and -

Daniel: Right, no, no, but I'm saying it doesn't work that way, that you don't give them both C's, you can't give them all C's, because 2 C's don't equal an A, I'm saying they can all have A's, in the same way that gravity is absolutely constant, what changes is mass, just like…what would change [in my story]

Lynn: Umm... I guess kind of what, now how I'm trying to think of it, now if we have something heavier going faster, then if we stacked up all those students together and a lot them only had C's, now all of a sudden they would all have an A, but truly, whether we pack them all together or drop them separately, they're C's, and when we drop something (*inaudible*) all we've essentially done is packed them all together, but a nickel's worth of weight, (*inaudible*) and if there are a hundred nickels of weight in something, it's still going to be the same gravity, cause they're each working on that little piece, like the nickel, like an economy of scale [] you still need the same amount for each little bit of mass, but on the other hand, you're not penalized for being heavier, because…

Daniel: Alright, if you have a roll of quarters and you drop it and film it and time how long it took, and then you have all these quarters strung out apart from each other,

Lynn: They come down the same

Daniel: Right. That's not an explanation, but it shows your principle.

In Daniel's metaphor of gravity as letter-grade, gravity is reconceived as an intensive property of an object that is not shared or enhanced by adding more objects or dividing an object into parts. By changing the metaphor, Daniel is able to reconcile attributing spatial location and quantification to gravity without needing to conserve it.

Daniel ends this conversation by suggesting an experiment that would make Lynn's reasoning very explicit: how the roll of quarters (heavy object, consolidated) would fall compared to having the quarters strung out (heavy object, separated into parts). Lynn and Daniel agree that they will come down the same.

***Summary and Discussion for Episode 1:*** Overall, Daniel's and Lynn's language in this first episode suggests the coordination of multiple substance-based ontologies of gravity, e.g., as stuff that comes in finite, quantifiable amounts within falling objects, or as an agentive substance that can "pull" or "work on" falling objects.

Daniel objects to Lynn's reasoning on ontological grounds, challenging her reluctance to conceptualize the amount of gravity as add-able or divisible. His objections highlight subtle differences in the way they are thinking about gravity. Daniel is thinking of gravity as quantifiable and add-able (i.e., conserved). Lynn, on the other hand, also quantifies gravity, but mostly rejects the idea of gravity as add-able, perhaps because she leans more heavily on the *force as agent* metaphor. Daniel's final analogy, which seems to connect with Lynn's ideas, compares gravity to a letter-grade. The final reconciled analogy taken up by both Lynn and Daniel conceptualizes gravity as an intensive property of an object — a property that is not enhanced or diluted by adding more objects or dividing an object into parts. Overall, their argument is reminiscent of what Brookes and Etkina call "ontological disambiguation"; Daniel and Lynn are trying to figure out what kind of entity gravity is by arguing about what attributes apply in this situation.

From the perspective of evaluating correctness, all the metaphors and analogies employed by Daniel and Lynn are ontologically incorrect, in that they are substance-based and/or they fail to disambiguate forces from fields [65]. So, as mentioned above, the instructors faced a tension between supporting the reflective argumentation between Lynn and Daniel as they struggle to make sense of a physical phenomenon and nudging Lynn and Daniel toward correct ontologies. Ultimately, as shown in the next episode, Daniel and another teacher from a different small group are able to build on the explanation developed by Lynn and Daniel to make significant progress toward reaching a correct explanation for the phenomenon.

**V. EPISODE 2: THE WHOLE GROUP BUILDS ON LYNN'S PRESENTED IDEA**

As usual in this workshop, the small-group discussion just analyzed was followed by a whole-group discussion for the purpose of sharing ideas between groups. In this section, we discuss how ideas presented by Lynn's small group were taken up by the larger group and to show how teachers come to construct the correct physics explanation for why objects of different masses fall with the same acceleration. The discussion, at times, focuses on the explanation from Lynn's and Daniel's group. When the discussion goes in other directions, we summarize the discussion here rather than presenting it in detail.

Andy opens the discussion with a bid for groups to contribute their thoughts:

Andy: The last thing we did was try to resolve in our little groups why it wasn't the case that gravity would grab the heavier bottle and pull it down more quickly than the other bottles? And I want to hear what thoughts you have on that.
(*Lynn raises her hand*)
Andy: Lynn?

Lynn: Well we came to the counterintuitive conclusion that gravity works on all things equally, regardless of their weight, and then we recalled that famous experiment where they dropped things and they arrived at the same time, or when Dave was telling us about dropping two items in a vacuum, that when it's just gravity you're dealing with, it's the same.
Andy: Then it doesn't care, as it were...
Lynn: Then it doesn't care what it weighs.
Andy: Okay. So that's a conclusion. Do you have a causal story for why that's so?
Lynn: No, I don't. I have all that evidence, but truly it's counterintuitive to me. I keep thinking that something that's heavier should fall faster. I don't understand the whole…

Lynn reiterates her sense, also stated at the start of the small-group discussion, that the correct answer is "counterintuitive," indicating that this counter-intuitiveness had not been fully resolved for her. Also note that Andy's utterance, "gravity would grab the heavier bottle," as well as Lynn's utterance, "gravity works on all things" suggest the metaphor of gravity as an agent. The strong anthropic sense in Andy's wording of the metaphor is striking. Andy was aware that the participants had been using such language in their small group discussions (as in Episode 1). As a facilitator, Andy's goal was to keep the conversation focused on meaning making in the teachers' own terms, and this included talking in terms that make everyday sense, even at the cost of technical precision or ontological correctness.

The class discussion at this point turns to other groups' ideas. Another teacher, Sam brings up the idea that the trajectory of the objects would depend on many factors such as its initial momentum, the weight, the shape of the object, and the height from which it is dropped. For about 20 minutes the whole group talks about how these different factors might play a role in determining the time of fall for an object. Andy lets the discussion take its course, occasionally asking participants to explain their causal story for how some condition or property makes an object fall faster or slower than another. After about 20 minutes, Andy makes a bid to bring the group back to the question of a heavier versus a lighter object:

Andy: So, there is an issue here, even if we know the answer, there's still the intuition that it's getting pulled down harder. So any ideas for reconciling that?
Marissa: I think it falls at the same rate of speed, no matter how much it weighs.
Andy: So everything falls at the same rate of speed, no matter how much it weighs. Right, so that's, um, and then some people have been saying that once you get to a certain height and or speed, you start to get away from that, but at least for the simple situation we're doing here, so that, that, does that explain why it happens? ... It very nicely clarifies what needs to be explained though, so that's important. What needs to be explained is, why do these fall at the same rate, even though one feels to me like it's getting pulled down harder?

At this point, Lynn rejoins the conversation, articulating some of the ideas and analogies that had come up in their small group discussion (Episode 1):

Lynn: Well we were really struggling with the germ of an idea, and we're not totally sure, but, we sort of have the idea that, like if you take a roll of quarters and one quarter, we know that if you drop them, they're going to land at the same time now. And to try to reconcile that, I was trying, we had a couple of analogies, I was trying to think of gravity as a constant force all over the earth, and in order to pull that one quarter down, it takes like two ounces, or two whatever, units of gravity to pull it down. So, when you just lump them all together, it doesn't change the fact that each one of those quarters in that roll is still going to need its little two whatevers of gravity, so putting them together doesn't make it any harder or any easier for it to come down, if each little part still needs that little bit of gravity, so if you thought about that coin roll, and slowly started separating them, just the fact that you're pushing them together doesn't, shouldn't really affect the fact that each need their own little bit of gravity, and we thought about a car as well, We said maybe it takes fifty gallons to move one of those semi trucks a mile. Well, when you add another semi truck to it, it doesn't make it take less energy, for the new space that's been added on, it needs more gasoline, so we kind of think of gravity as this is something that each little amount of space needs a certain amount of to be pulled down, and it doesn't matter if you lump the space together, you pull them apart, it's all going to work the same, if you think of that roll of quarters as just a mess of little quarters put together,

Here Lynn analogizes gravity to a constant force distributed over the earth; but the sense of constancy she expresses is not of an absolute constancy, but constant for "each little part." Her analogy of gravity to fuel makes this clearer. She says that if one truck needs 50 gallons to move a mile, then two trucks should take twice as much to move a mile; one cannot add another truck and expect the same 50 gallons to move both trucks by a mile. In that sense, she is thinking of the fuel being constant *per truck* and making sense of how gravity could be constant "for each little amount of space." In her quantification of gravity as "2 whatevers," in phrases such as "little bit of gravity," and her analogy of gravity as fuel, Lynn is ontologically treating gravity as a material substance. And she is doing so productively, as it is beginning to help her make progress in understanding why heavier things do not fall faster.

Next, Andy teases apart the various details to Lynn's argument in order to really understand it and to make sure that the class has a sense of Lynn's explanation of gravity:

Andy: They just happen to be together
Lynn: They happen to be stuck together, but they're still going to be pulled down at the same rate.
Andy: Let me see if I can understand that argument. So you're saying, we'll start with the quarters, and then we'll talk about the car.
Lynn: Okay
Andy: So you're saying, here's a quarter (drawing on board), and it /felt/ some little bit of gravity to pull it down, 2 ounces of gravity
Lynn: I don't know what gravity's measured in *(chuckling)*
Andy: 2 ounces of gravity pulls down the one quarter...If it's a bunch of quarters, well, each one needs its own 2 ounces, and they all have the gravity downward pull, and then the next step is, okay, instead of there actually being air between the quarters, they're pressed together into a roll. And you're saying that's a difference that is not really a difference. The point is, it's still that number of quarters with gravity pulling on each one however hard it pulls it, and if there's, what, 8 quarters instead of 1, you've got 8 2-ounce pulls instead of 1 2-ounce pull.

Here, as earlier, Andy responds using the same kind of language that the teacher-participants are using, reflecting the same *matter* and *agency* attributes in phrases such as "2 ounces of gravity" and "gravity pulling on each." Such revoicing, we feel, helped maintain the framing of the discussion as an activity to make sense of one's own ideas.

Another teacher, Nancy, raised the argument that lumping the quarters together versus having air between them would affect how they fall due to the added air resistance in the latter case. Andy responded to that argument by saying that air resistance would indeed play a role if we were thinking about dropping feathers lumped together or separated out, but not for pennies since they are so light and compact. Sam brought up the idea of how conveyor belts on airports move at a constant speed irrespective of the weight they carry. But when probed, he did not offer a mechanism for how that happens, just that gravity could be like a conveyor belt that maintains the objects at a constant speed. After this, Andy brings the discussion back to Lynn's argument:

Andy: In this story, gravity is adjustable in that the single quarter feels one unit of pull, the 8 pack of quarters feels 8 units of pull, so by this story, it's like, it would be like the walkway knowing to start working harder somehow, and the way gravity knows is that it just pulls each quarter separately.

Note that this is exactly the point that Daniel was contending in the small group discussion. Daniel wanted the addition and subtraction of gravity with added or removed mass as an allowable move, while Lynn at that time was rejecting this idea since she was focusing on the idea of *sameness* of gravity per unit object. Here Lynn does not object to Andy's move of adding the gravity for the eight quarters – perhaps due to Andy's status as an instructor for the workshop. Another speculation is that Lynn has warmed up to the notion of gravity increasing proportionally as you add objects. Andy's articulation of which object feels what amount of pull ("if there's, what, 8 quarters instead of 1, you've got 8 2-ounce pulls instead of 1 2-ounce pull, might also have lent clarity to what is being added; as discussed earlier, the ambiguous use of pronouns might have contributed to Daniel's and Lynn's disagreement about the add-ability of gravity.

Now Daniel – who earlier has been arguing for being able to add gravity when joining together multiple lumps of mass– takes on Andy's idea of lumping the quarters asking if the inertia of the quarters would also be proportional to the number of quarters:

Daniel: Well, the 8 quarters also have 8 times the inertia, right? So it's going to, if they're heavier, they're going to have more mass, so they have more inertia, um, does that factor in?
Andy: So, more inertia. So what is inertia meaning for us, you right now?
Daniel: Um, it's going to have, uh, when you drop it, it wants to stay at rest, but gravity's pulling it down, so it's got to overcome that willingness to stay at rest. The more massive it is, the more it's going to want to stay, to not move.
Andy: So you're saying things that are just sitting somewhere don't want to move, and you're saying the, uh, the bigger heavier thing has more "not want to moviness" to it then a lighter thing, so.
Daniel: 8 quarters stacked together is 8 more times not willing to move than one quarter.
Andy: So the 8 quarters is 8 times as hard to move.
Dave: But that means there's 8 times as much gravity pulling it down.
Daniel: Right. And that's why they fall at the same /time/.
Andy: But there's 8 times as much gravity pulling it down. (overlapping with Daniel's utterance)

Daniel brings an idea into the discussion, inertia, that he clarifies as "willingness to stay at rest" and that the 8

quarters put together will be 8 times more difficult to move. Dave reiterates the idea that the 8 quarters also have 8 times the pull on a single quarter. This idea later forms the bridge to arriving at the canonically correct explanation.

A few aspects of Daniel's reasoning are worth noting. First, his description of inertia in colloquial language is powerful in conveying the abstract idea of inertia. It also suggests that Daniel is taking an epistemological stance in which he expects that abstract science concepts such as inertia can connect to everyday thinking, a stance that the workshop specifically targeted. Further, Dave's and Daniel's language of "gravity pulling it down" continues to reflect the sense of *gravity as an agent* that has been present in the discussion from the start.

At this point in the discussion several people start speaking. Andy ties together the two ideas of gravitational force and inertia increasing proportionally with increased mass.

Andy: So in this story now, there's, uh, we keep getting these battles, maybe it's just my male GI Joe thing. There's this battle, on the one hand, gravity wants to move the thing when you drop it, on the other hand, something doesn't want to get moved. If there weren't gravity and I let go of this thing, what would it do.

Several voices: (*inaudible*)

Andy: Yeah, it doesn't want to move, This garbage can's not going to move unless something kicks it. Um, so by this story, there's this battle between the pull of gravity and the tendency of the thing not to want to move. And for a single quarter, there's a certain amount of "not wanting to moviness" and a certain amount of pull, and the result of that battle leads to whatever fall you see. For 8 quarters, how much more "not wanting to moviness" does it have?

Voices: 8 times

Andy: It's 8 times as reluctant to move, but...

Voices: 8 times the...

Andy: It's got 8 times as much pull on it.

Erwin: That makes sense.

Andy: With the anvil, it's huge, it's 500 times more reluctant to move, but it's got 500 times more pull on it, so it cancels each other out.

In this final segment, Andy tried to tie together the ideas brought by Daniel and Dave, that the higher gravitational pull on an object of larger mass is exactly compensated by its higher inertia, so that objects of different masses have the same acceleration due to gravity. The leading nature of his facilitation at this point, however, makes it difficult to gauge how well the teacher participants are "getting it."

In the next section we continue our discussion of how the material substance ontology of gravity was productive for the group in making progress toward this correct explanation.

### VI. DISCUSSION AND IMPLICATIONS

We presented a case study of in-service middle- and elementary-school teachers, in a professional development setting, trying to figure out a causal explanation for why dropped objects of different masses fall equal distance in equal time. Four teachers discussing this within their small group acknowledge that they know the phenomenon through standard physics demonstrations but are puzzled about an explanation, since it feels as if the more massive object is pulled down harder. We analyzed this discussion because our attention was drawn to the teachers' ontologically-rich analogies and reasoning. Upon closer analysis, we found that most of the utterances encoded an ontology of *gravity as an agent* and *gravity as a material substance* – ideas which have often been marked by science education researchers as "mis-ontologies" that represent misconceptions and hinder conceptual development. We argue that these "mis-ontologies" were productive in leading teachers to construct a correct explanation for the phenomenon.

The idea that gravity acts on each part of an object as if that part were not attached to the rest of the object, and hence the (heavy) object falls in the same way as its (lighter) pieces would fall, was used by Galileo [66]. For the teachers, this reasoning flowed directly from their thinking of different pieces (quarters, cars, parts of a tootsie roll) *having* a certain amount of gravity pulling them down (coordinating two ontological resources: *gravity as substance* contained by falling objects and *gravity as a material agent* acting on falling objects). In other words, the Galilean reasoning emerged because of, not in spite of, the teachers' "mis-ontologies" of gravity. And this idea of each coin in a roll of coins feeling a certain amount of gravity then fed into the Newtonian compensation argument whereby the heavier object feels more gravitational pull but also puts up more resistance to getting moved. So, in the discussion, the mis-ontologies of gravity propagated all the way to this Newtonian argument — and later in the session, a discussion of Newton's second law. To be clear, we do not dispute that the teachers would need to think about forces and gravity in some ontologically different ways in order to achieve an expert understanding of those concepts and of Newtonian mechanics more generally. Our point is simply that the teachers made striking conceptual progress reasoning in terms of "incorrect" ontologies of force and gravity.

More importantly, this episode suggests that in some cases, attempts to stamp out novices' substance-based ontologies might be counter-productive for their learning because such interventions would disrupt participants' sense-making about the physical phenomenon—their construction of causal, coherent explanations consistent with the evidence. Having the space to value and make sense of their own ideas helped Lynn and Daniel add greater clarity to their arguments. An early intervention by the facilitator stating the "correct" language to use could have directed attention away from their own arguments and towards using the "right" words.

### A. Addressing counterarguments

We want to start by addressing counter-arguments that could be posed by researchers who have argued that substance-based ontologies for non-material entities lead to robust misconceptions [17–20,23,28,31–40,61] and hence that instructors should nudge learners away from these mis-ontologies. One such argument is that the teachers in our group, despite engaging in extensive discussion, did not change their ontological view of gravity. Indeed, the mis-ontology might have been reinforced by the lack of attention drawn to it by the facilitator who, at times, co-opted the language encoding the incorrect ontology. By this argument, the current episode shows how, without appropriate intervention, ontological misconceptions do not simply get corrected via discussion.

On this point, we agree: the teachers did not undergo any wholesale ontological change within this episode.

The data, however, challenge the instructional implication that incorrect ontological classification of science concepts is categorically unproductive and should be stamped out as soon as possible [21]. More importantly, this episode suggests that in some cases, attempts to stamp out novices' substance-based ontologies might be counter-productive for learning because such interventions would disrupt participants' sense-making about the physical phenomenon—their construction of causal, coherent explanations consistent with the evidence.

Another counterargument is that this episode is too unusual to inform the debate about how to address students' naïve ontologies, since in most usual cases, learners using incorrect ontologies will not make such conceptual progress. By this argument, our data may show that it's *possible* for mis-ontologies to support productive reasoning but not that it's common. We have two responses to this. First, while acknowledging that this episode was unusual in how quickly and articulately the small group came to a "cool" (in this case, Galilean) explanation, we note that in our summer workshops, teachers commonly relied on "incorrect" ontologies to make conceptual progress. The very next summer, for instance, another teacher (supported by a facilitator) came up with a gravity-inertia "compensation" explanation for why heavy and light objects fall together, without changing her agent-like ontology of gravity. And in multiple summers, teachers have conceptualized electric current as "stuff" to support productive reasoning about what happens to the current at a junction in a circuit. (In Chi et al's framework, it is incorrect to think of current in substance-like ways.) So, again, the episodes presented in this paper are unusually dramatic but not unusual in challenging the view that mis-ontologies necessarily block conceptual progress. Second, we note that even in most reform-oriented physics curricula, due to time constraints, students are guided to make fairly "direct" conceptual progress toward expert conceptions and ontologies. In those contexts, allowing mis-ontologies to persist for even a short time could be inefficient for the vast majority of students, and hence studies conducted in those contexts are likely to show that mis-ontologies are a hindrance to learning. In our summer workshops, by contrast, we are not trying to reach pre-determined conceptual goals within a fixed time and hence the teachers can take more "indirect" paths toward improved conceptual understandings. This learning environment allows us to see the potential productivity of reasoning grounded in "incorrect" ontologies more clearly than is evident in more guided, time-constrained environments. Our point is that when mis-ontologies act as a barrier to learning, the barrier arises not from a fundamental cognitive limitation on building conceptual progress out of "incorrect" ontologies, but rather, from the incompatibility of those mis-ontologies with the particular conceptual pathways along which PER-based curricula guide students.

### B. Instructional implications and educational values

We started the paper by summarizing a debate on novice cognition, and we want to end with a note on educational values. In her book, *The Having of Wonderful Ideas*, Duckworth decries the notion that "there is one best way to understand" [67] (p. xii):

> The essays in this book start instead from the premise mentioned above – that there is a vast array of very different adequate ways that people come to their understanding.

    Curriculum, assessment, teacher education programs – and all of our teaching – must seek out, acknowledge, and take advantage of the diversity of ways that people might take toward understanding. We cannot plan "*the logical sequence*" through a set of ideas, especially if we want schools to make sense for students whose backgrounds differ from out own….we must find ways to present subject matter that will enable learners to get at their own thoughts about it. Then we must take those thoughts seriously, and set about helping students to pursue them in greater breadth and depth." ( [67] p. xiii) [emphasis and paragraph break as in original]

In step with Eleanor Duckworth, we feel that classrooms – and professional development spaces – should be a place for the "having of wonderful ideas," that learners should feel that the classroom is a place where they are free to explore and play with ideas, make mistakes, and reflect on their progress, rather than a place that undervalues their experiences and ideas and defines the purpose of learning as faithfully absorbing the ideas generated by others (scientists, in science classrooms) in a prescribed manner. A science learning environment should be a place for discovery rather than a place *only* for learning of the end products of other people's discovery. To be clear, we are not saying that there is no place for learning knowledge produced by communities of scientists and mathematicians. But increased emphasis on "getting" the correct conceptions (and correct ontologies for those concepts) and recommendations to strongly guard against any incorrectness (in terms of conceptual knowledge, terminology, ontology) are a threat to the classroom as a place for pursuing the authentic *practices* of science. And we feel that with respect to this need the professional development space for teachers should also provide teachers experiences with the authentic pursuit of science as an exploration into natural phenomenon, and creating causal, coherent explanations for such phenomenon [68,69]. How can we expect teachers who have not undergone such experiences themselves to create such experiences for their students?

    The experience of struggling with trying to explain how objects of different masses fall with the same acceleration was a powerful experience for Lynn – one that she refers back to many times over the course of 4 years in our professional development program. She always remembers it fondly and with excitement as helping her experience the excitement of doing science. In an interview during the fourth year into the project (with a team member who joined in year 2 and wasn't deeply involved in the professional development part of the project), Lynn says:

> On a personal level, the first year I did this and we did the, one of the things where you drop book and feather, which hits first? And we had to figure out for ourselves how that occurs? What happens? And I, it was very frustrating working through that, but when I actually figured it out for myself, that was probably one of the most exhilarating intellectual moments I've ever had in my life. It was just really astonishing. And then, that was cool because that happened with me that summer, and then I started using it in classroom, and I can see same kind of epiphany occurring with students, and I know how exhilarating and empowering it is to have that kind of experience. So that's probably been the most amazing part.

Lynn added that the "non-judgmental" atmosphere made her and others feel "they have something really significant to contribute. And that's been… amazing… for me, just in life in general."

An early focus on correcting Lynn's ontology and driving her to think about gravity as a force, and hence as an interaction – not as a material agent/substance – would most likely have robbed Lynn of this transformational experience. And in our vision of professional development programs for science teachers, we value such experiences more than creating conceptual change.

## ACKNOWLEDGEMENTS

We thank Colleen Gillespie for the transcription of the video data. We thank the teachers for participating in our workshop and allowing us to videotape their discussions. We thank Amy Green for conducting interviews with teachers. We thank David Hammer, Colleen Gillespie, and Kweli Powell for discussions about this analysis. This work was supported by funding from NSF MSP 0831970.